\documentclass[aip,jcp,reprint]{revtex4-1}

\usepackage{amsmath}
\usepackage{array}
\usepackage{makecell}

\usepackage{graphicx}
\usepackage{dcolumn}
\usepackage{bm}
\usepackage{color}

\usepackage[utf8]{inputenc}
\usepackage[T1]{fontenc}
\usepackage{mathptmx}
\usepackage{etoolbox}

\makeatletter
\def\@email#1#2{%
 \endgroup
 \patchcmd{\titleblock@produce}
  {\frontmatter@RRAPformat}
  {\frontmatter@RRAPformat{\produce@RRAP{*#1\href{mailto:#2}{#2}}}\frontmatter@RRAPformat}
  {}{}
}%
\makeatother

\newcommand{\beq}{\begin{equation}\begin{aligned}}
\newcommand{\eeq}{\end{aligned}\end{equation}}

\begin{document}

\preprint{AIP/123-QED}

\title{Orbital-Optimized Quasi-Variational Distinguishable Cluster Doubles Method}

\author{Fangcheng Wu}
\email{fc.wu@fkf.mpg.de}
\affiliation{Max Planck Institute for Solid State Research, Stuttgart, Germany}

\author{Daniel Kats}
\email{d.kats@fkf.mpg.de}
\affiliation{Max Planck Institute for Solid State Research, Stuttgart, Germany}

\date{\today}

\begin{abstract}
We present the orbital-optimized quasi-variational distinguishable cluster doubles (OQVDCD) 
method, obtained by applying the distinguishable cluster (DC) approximation
to orbital-optimized quasi-variational coupled cluster doubles (OQVCCD). 
The resulting method retains a Hermitian-like energy functional, is size-extensive, 
obeys the generalized Hellmann-Feynman theorem at a stationary point,
is exact in the isolated two-electron and two-hole limits, and scales as $\mathcal{O}(o^2v^4)$,
like coupled cluster singles and doubles (CCSD). For the tested closed-shell reaction energies,
it is shown that OQVDCD reduces the mean absolute deviation relative to OQVCCD 
from 1.20 to 0.81 kcal/mol, bringing the errors close to distinguishable cluster singles and
doubles (DCSD).
In the strongly correlated examples considered here, OQVDCD shifts the OQVCCD energies
toward the benchmark, although the convergence remains somewhat less robust than for DCSD.
\end{abstract}

\maketitle

\section{Introduction}

Ab initio wavefunction methods provide a systematic route to accurate descriptions and 
predictions of electronic structure.
In conventional single-reference methods, one starts from the determinant obtained from
Hartree-Fock (HF) self-consistent field (SCF) theory and then adds correlation corrections
to that reference wavefunction.\cite{Fock1930}
Established post-HF approaches include Møller–Plesset perturbation theory (MP),\cite{MP1934}
configuration interaction (CI),\cite{Loewdin1955} and traditional coupled cluster (TCC) theory.\cite{Cizek1966}

TCC owes much of its success to its size-extensivity, its systematic approach to full
CI (FCI) with increasing excitation rank, and its ability to yield chemically accurate
relative energies for many weakly correlated systems already at the coupled cluster
singles, doubles, and perturbative triples [CCSD(T)] level.\cite{CCSD_T}
However, truncated TCC methods are not variational and usually fail in the presence of
strong static correlation, where a single-determinant reference is no longer adequate.
This failure commonly appears as unphysical maxima in potential-energy curves and
collapse below the FCI energy.\cite{Knowles2010}
Multireference (MR) methods have therefore been developed to treat static correlation more reliably,
including state-specific MRCC\cite{Mukherjee1998, Koehn2013} and internally contracted MRCC.\cite{jiri1969, Evangelista2011, Koehn2011}
Within the single reference framework, improvements include spin-flip,\cite{krylov2001}
active space,\cite{Head-Gordon1998, Piecuch2010} and renormalization.\cite{Piecuch2000}
There are also alternative single-reference theories with an exponential ansatz,
such as variational coupled cluster (VCC),\cite{vcc-benchmark}
extended coupled cluster (ECC),\cite{ecc-1983, ecc-1987,ecc-2006-Piecuch} and
unitary coupled cluster (UCC),\cite{ucc1,ucc2} that improve on TCC at the same truncation level,
but at a substantially higher computational cost.\cite{Evangelista2011-Alter}

One natural starting point for such alternatives is VCC, which tends to yield robust behavior in strongly correlated regimes, thanks to its
variational character.\cite{Knowles2010,VanVoorhis2000}
VCC evaluates the energy expectation value of an exponential wavefunction ansatz and
determines the amplitudes by minimizing the functional. Some effort has therefore gone
into approximations that retain the size-extensivity with some of the variational features,\cite{KnowleszLPFD2010, KnowlesAVCCD2011, XCC1988, ICC1991}
including the quasi-variational coupled cluster (QVCC) family proposed by Knowles and co-workers.\cite{QVCC-J2, QVCC-J3, QVCC-J4, QVCC-J5, QVCC-J6, QVCC-J7, QVCC-J8}
QVCC doubles (QVCCD) is the foundation of this framework.
Rather than simply truncating the infinite VCCD energy expansion, QVCCD partially retains
higher-order doubles contributions in a finite differentiable functional.
Because the retained terms remain linked, QVCCD is rigorously size-extensive.
Orbital relaxation enters through orbital optimization, and the resulting OQVCCD retains the $\mathcal{O}(o^2v^4)$ scaling
of coupled cluster singles and doubles (CCSD).
In practice, QVCC shows more upper-bound character than TCC and behaves more physically
in strongly correlated regimes.

An alternative approach to improve on TCC is to modify the amplitude equations by excluding or altering
selected contributions.
Examples include coupled-electron pair approximations (CEPA),\cite{CEPA-1-Meyer, CEPA-2-Neese}
approximate coupled pairs with quadruples (ACPQ),\cite{Paldus1984}
$n$-electron-exact coupled cluster ($n$CC),\cite{Bartlett2006,Bartlett2007}
parameterized coupled cluster singles and doubles (pCCSD),\cite{pccsd-Huntington2010, pccsd-Huntington2012}
and the distinguishable cluster (DC) modification.\cite{D1-Kats2013,D2-Kats2014,D3-Kats2015,D4-Kats2016,D5-Kats2018,D6-Kats2019,D7-Kats2019,D8-Kats2021,D10-Kats2024,D11-Kats2025,D12-Kats2025}
DC often performs well across a wide range of systems.\cite{Rishi2019}
In the CC language, the central idea is to remove selected quadratic exchange terms
in which electrons originating from different double-excitation clusters are exchanged,
while keeping the spin-orbital amplitudes antisymmetric. DCSD improves upon CCSD not only
in strongly correlated cases but also for many near-equilibrium problems dominated by
dynamic correlation.
DC modification can substantially improve reaction-energy predictions while preserving
the computational scaling and the formal properties of the parent TCC truncated method.

Although DC modification has shown strong performance in TCC,
the resulting method inherits the non-Hermitian character of the parent method.
One motivation for extending it to a single-reference method with a Hermitian-like
energy functional is its potential use in embedding schemes with lower-order correlation
treatments for larger systems.\cite{christlmaierFull2022}
Here we apply the DC idea to OQVCCD and introduce orbital-optimized quasi-variational
distinguishable cluster doubles (OQVDCD).
The aim is to retain the formal advantages of the quasi-variational framework while improving
the description of dynamically correlated problems and assessing how much of that improvement
carries into more strongly correlated tests.

\section{Theory}
\subsection{TCC and VCC}
TCC and VCC share the exponential wavefunction ansatz in which the wavefunction $\Psi$ 
is parameterized by the exponential excitation operator $e^{\hat{T}}$ acting on 
a Slater-determinant reference $\vert 0\rangle$,
\beq
\vert \Psi \rangle = e^{\hat{T}}\vert 0\rangle,
\eeq
where the cluster operator $\hat{T}$, the single-excitation operator $\hat{T}_1$,
and the double-excitation operator $\hat{T}_2$ are defined as
\beq
\hat{T} = \sum_{n} \hat{T}_n
\eeq
\beq
\hat{T}_1 = t^{i}_{a}\hat{a}^\dagger_a\hat{a}_i  
\eeq
\beq
\hat{T}_2 = \frac{1}{4}t^{ij}_{ab}\hat{a}^\dagger_a\hat{a}^\dagger_b\hat{a}_j\hat{a}_i
\eeq
with $n$ the excitation rank and $\hat{T}_n$ the $n$-fold excitation cluster operators.
The Einstein summation convention is used throughout.
The quantities $t$ are excitation amplitudes, and $\hat{a}^\dagger$ and $\hat{a}$ are
creation and annihilation operators.
The indices $a,b,\ldots$ denote virtual spin orbitals, whereas $i,j,\ldots$ denote occupied spin orbitals.
Inserting the exponential ansatz into the time-independent Schrödinger equation,
\beq
\hat{H} e^{\hat{T}}\vert 0\rangle = E e^{\hat{T}}\vert 0\rangle,
\eeq
two different routes can be used to obtain the energy: projection or variation, depending on whether a TCC or VCC formulation is adopted.

In TCC, usually formulated using the similarity-transformed Hamiltonian
$\bar{H} = e^{-\hat{T}}\hat{H} e^{\hat{T}}$, the Schrödinger equation can be rewritten as
\beq
e^{-\hat{T}}\hat{H} e^{\hat{T}}\vert 0\rangle = E \vert 0\rangle,
\eeq
so the energy is obtained by projecting this expression onto the reference determinant $\langle 0\vert$,
\beq
\langle 0\vert e^{-\hat{T}}\hat{H} e^{\hat{T}}\vert 0\rangle = \langle e^{-\hat{T}}\hat{H} e^{\hat{T}} \rangle = E_{TCC},
\eeq
where $\langle\hat{O}\rangle$ denotes the expectation value with respect to the reference determinant.
Furthermore, the Baker-Campbell-Hausdorff expansion of $\bar{H}$ truncates naturally after
the quartic term, which is one source of the favorable computational cost of TCC,
\beq
\bar{H} = \hat{H} + [\hat{H},\hat{T}] + \frac{1}{2!} [[\hat{H},\hat{T}], \hat{T}]+...
\eeq
The amplitudes $t$ are determined from projected amplitude equations.
If we restrict the cluster operator to doubles, and 
project the similarity-transformed Schrödinger equation onto the doubly substituted determinants
$\langle \Phi^{ij}_{ab}\vert$, only terms up to quadratic order remain because higher terms
do not connect to the doubles manifold,
\beq
\langle \Phi^{ij}_{ab}\vert \hat{H}_N + (\hat{H}_N\hat{T}_2)_C + \frac{1}{2}(\hat{H}_N\hat{T}^2_2)_C\vert0\rangle = 0,
\eeq
where $\hat{H}_N = \hat{H}-\langle0\vert \hat{H}\vert0\rangle$ and $C$ denotes the connected terms.

\begin{figure}
\begin{minipage}{0.24\linewidth}
\centering
\includegraphics[width=\linewidth]{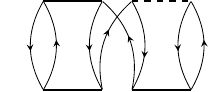}\\
\textbf{~~A}
\end{minipage}
\begin{minipage}{0.24\linewidth}
\centering
\includegraphics[width=\linewidth]{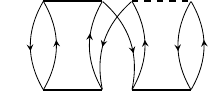}\\
\textbf{~~B}
\end{minipage}
\begin{minipage}{0.24\linewidth}
\centering
\includegraphics[width=\linewidth]{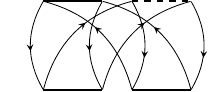}\\
\textbf{~~C}
\end{minipage}
\begin{minipage}{0.24\linewidth}
\centering
\includegraphics[width=\linewidth]{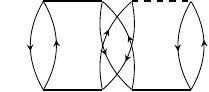}\\
\textbf{~~D}
\end{minipage}
\caption{Diagrammatic representation of $\langle\hat{T}_2^\dagger \hat{H}_N\hat{T}_2^2\rangle_L$
 terms in VCCD energy functional}
\label{fig:dc_diagrams}
\end{figure}

In VCC, by contrast, one evaluates the energy expectation value.
Because the unlinked terms in the numerator cancel against the denominator
$\langle e^{\hat{T}^\dagger} e^{\hat{T}}\rangle$, the resulting expression is
an infinite expansion in linked terms,
\beq
E_{VCC} =& \frac{\langle e^{\hat{T}^\dagger}\hat{H} e^{\hat{T}}\rangle}{\langle e^{\hat{T}^\dagger} e^{\hat{T}}\rangle} = \langle \hat{H}\rangle+\langle e^{\hat{T}^\dagger}\hat{H}_N e^{\hat{T}}\rangle_L\\
=& \langle \hat{H}\rangle+2\langle\hat{H}_N\hat{T}\rangle_L+\langle\hat{T}^\dagger \hat{H}_N\hat{T}\rangle_L \\ 
&+ 2\cdot \frac{1}{2!}\langle\hat{T}^\dagger \hat{H}_N\hat{T}^2\rangle_L  + \frac{1}{2!}\cdot \frac{1}{2!}\langle\hat{T}^{2\dagger} \hat{H}_N\hat{T}^2\rangle_L\\
&+\dots,
\label{eq:vcc_expansion}
\eeq
and the amplitudes are determined by variational optimization of the functional with respect to $t$,
\beq
\frac{\partial E_{VCC}}{\partial t} = 0.
\eeq
In the VCCD energy functional, the quadruple excitation contribution
$\langle\hat{T}_2^\dagger \hat{H}_N\hat{T}_2^2\rangle_L$ can be separated into four terms,
labeled A, B, C, and D from the first through fourth lines of Eq.~(\ref{eq:VCCD_quadruple_terms}),
with antisymmetrized Goldstone diagram (ASG) representations shown in Fig.~\ref{fig:dc_diagrams}.
\beq
\langle\hat{T}_2^\dagger \hat{H}_N\hat{T}_2^2\rangle_L =& \frac{1}{4}(t^\dagger)^{ab}_{ij}[-(1-\tau_{ab})(v_{kl}^{cd}-v_{lk}^{cd}) t_{ca}^{kl}t_{db}^{ij}\\
&-(1-\tau_{ij})(v_{kl}^{cd}-v_{lk}^{cd}) t_{cd}^{ki}t_{ab}^{lj}\\
&+\frac{1}{2}(v_{kl}^{cd}-v_{lk}^{cd}) t_{cd}^{ij}t_{ab}^{kl}\\
&+(1-\tau_{ab})(1-\tau_{ij})(v_{kl}^{cd}-v_{lk}^{cd}) t_{ac}^{ik}t_{bd}^{jl}],\\
\label{eq:VCCD_quadruple_terms}
\eeq
Here $v_{pq}^{rs}$ denotes the two-electron integrals $\langle pq|rs \rangle$,
and $\tau_{ab}$ and $\tau_{ij}$ are permutation operators that exchange the indices of
the virtual and occupied pairs, respectively. 

\subsection{Overview of QVCCD}
Because VCC combines size-extensivity and variationality, it is natural to seek an approximation 
that preserves these properties (at least approximately) while retaining a computational
scaling competitive with that of TCC. QVCC is one such approximation and has several features
that make it attractive.

The QVCC framework is built on QVCCD,
to which orbital optimization and perturbative higher-excitation corrections can be added.
\cite{QVCC-J4, QVCC-J5, QVCC-J6, QVCC-J7}
The QVCC energy functional is closely related to linked pair functional doubles theory (LPFD)\cite{KnowleszLPFD2010}
and approximate variational coupled cluster theory (AVCC).\cite{KnowlesAVCCD2011}
These approaches retain the first three terms of the VCC expansion in Eq.~(\ref{eq:vcc_expansion})
and replace the doubles operator $\hat{T}_2$ in the first- and second-order contributions
by transformed doubles operators $\,\!{_1}\hat{T}_2$ and $\,\!{_2}\hat{T}_2$,
\beq
E = \langle \hat{H}\rangle + 2\langle\hat{H}~\!{_2}\hat{T}_2\rangle + \langle\,\!{_1}\hat{T}_2^\dagger\hat{H}_N~\,\!{_1}\hat{T}_2\rangle.
\label{eq:energy_functional}
\eeq
$\,\!{_1}\hat{T}_2$ and $\,\!{_2}\hat{T}_2$ are defined as double-excitation cluster operators
with transformed amplitudes,
\beq
\,\!{_q}\hat{T}_2=\frac{1}{4}\,\!{_q}t^{ij}_{ab}\hat{a}^\dagger_a\hat{a}^\dagger_b\hat{a}_j\hat{a}_i,
\eeq
in which $q=1$ or $2$. In QVCCD,
\beq
\,\!{_q}t^{ij}_{ab} &= 2 \cdot [\frac{1}{2}(1-\tau_{ab})(\,\!{_A}U^{-\frac{q}{2}})_a^ct^{ij}_{cb} ]\\
&+ 2 \cdot [\frac{1}{2}(1-\tau_{ij})(\,\!{_B}U^{-\frac{q}{2}})_k^it^{kj}_{ab}]\\
&-1\cdot[\frac{1}{2}(\,\!{_C}U^{-\frac{q}{2}})^{ij}_{kl}t^{kl}_{ab}]\\
&-2\cdot [\frac{1}{4}(1-\tau_{ab})(1-\tau_{ij})(\,\!{_D}U^{-\frac{q}{2}})^{ic}_{ak}t^{kj}_{cb}],\\
\eeq
with transformation matrices
\beq
\,\!{_A}U^a_b &= \delta^a_b + \frac{1}{2}(t^\dagger)_{ij}^{ac}t^{ij}_{bc} \\
\,\!{_B}U_i^j &= \delta_i^j + \frac{1}{2}(t^\dagger)_{ik}^{ab}t^{jk}_{ab} \\
\,\!{_C}U^{ij}_{kl}&=(\delta_k^i\delta_l^j-\delta_k^j\delta_l^i) +\frac{1}{2}(t^\dagger)^{ab}_{kl} t^{ij}_{ab}\\
\,\!{_D}U_{bi}^{ja} &= \delta_b^a\delta_i^j + (t^\dagger)_{ik}^{ac}t^{jk}_{bc}.\\
\eeq
Treating the four indices of $\,\!{_C}U$ and $\,\!{_D}U$ as pair indices,
the transformation matrices $\,\!{_A}U$, $\,\!{_B}U$, $\,\!{_C}U$, and $\,\!{_D}U$ are raised
to the power $-\frac{q}{2}$.
These matrices are defined from unit matrices plus reduced-density-matrix contributions.
They are positive definite, thus negative and noninteger powers of these matrices are well defined.\cite{QVCC-J2}

Since the VCC energy expansion is infinite, any simple truncation would destroy its methdological properties.
The transformed operators effectively reintroduce an infinite linked series through the inverse matrices.
The amplitudes are determined by differentiating the energy functional,
and solving the resulting stationary conditions.
The derivatives of the inverse matrices can be obtained with the standard Fréchet-derivative theory.\cite{Higham-functions_of_matrix}
Because the QVCCD functional is stationary with respect to the amplitudes,
it obeys a generalized Hellmann-Feynman relation,
although it is not a rigorous upper bound to the ground-state energy.

One design criterion for the transformed operators is the exactness with respect to FCI
in the isolated two-electron and two-hole limits.
In these limits, higher powers of the doubles operator in VCCD annihilate when applied to
the reference determinant, so
 $e^{\hat{T}_2}\vert0\rangle = (1+\hat{T}_2)\vert 0\rangle$,
 which means that VCCD reduces to FCI without single excitations or, equivalently, CI doubles (CID). 
\beq
E =& \frac{\langle (1+\hat{T}_2^\dagger)\hat{H}(1+\hat{T}_2)\rangle}{\langle (1+\hat{T}_2^\dagger)(1+\hat{T}_2)\rangle}\\
=&\langle \hat{H}\rangle+ \frac{2\langle\hat{H}\hat{T}_2\rangle +\langle\hat{T}_2^\dagger(\hat{H}-\langle \hat{H}\rangle)\hat{T}_2\rangle}{1+\langle\hat{T}_2^\dagger\hat{T}_2\rangle}.
\eeq
In QVCCD, the transformed cluster operators are designed to satisfy the following rule in these limits,
\beq
\,\!{_q}\hat{T}_2=\left(1+\langle \hat{T}_2^\dagger\hat{T}_2\rangle\right)^{-\frac{q}{2}}\hat{T}_2,
\eeq
so that the QVCCD energy functional is exact to VCCD and CID
in the isolated two-electron and two-hole systems.

\begin{figure}
\begin{minipage}{0.24\linewidth}
\centering
\includegraphics[width=\linewidth]{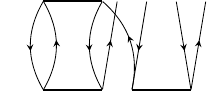}\\
\textbf{~~A}
\end{minipage}
\begin{minipage}{0.24\linewidth}
\centering
\includegraphics[width=\linewidth]{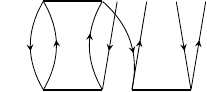}\\
\textbf{~~B}
\end{minipage}
\begin{minipage}{0.24\linewidth}
\centering
\includegraphics[width=\linewidth]{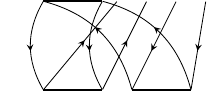}\\
\textbf{~~C}
\end{minipage}
\begin{minipage}{0.24\linewidth}
\centering
\includegraphics[width=\linewidth]{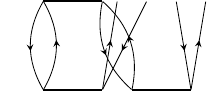}\\
\textbf{~~D}
\end{minipage}
\caption{Diagrammatic representation of $\mathcal{O}(t^3)$ terms of transformed double excitation operators $\,\!{_2}\hat{T}_2$}
\label{fig:dc_diagrams_open}
\end{figure}

A second design criterion is to retain all $\mathcal{O}(t^3)$ subsets of terms present in VCCD.
The transformation matrices are chosen so that QVCCD reproduces the exact quadruple contribution
through the VCC term $\langle\hat{T}_2^\dagger \hat{H}_N\hat{T}_2^2\rangle_L$
in Eq.~(\ref{eq:VCCD_quadruple_terms}),
\beq
2\langle\hat{H}(\,\!{_2}\hat{T}_2)\rangle =& 2 \cdot \frac{1}{4} (v^{ab}_{ij}-v^{ba}_{ij})\,\!{_2}t^{ij}_{ab}\\
=& 2\langle \hat{H}\hat{T}_2\rangle + \langle\hat{T}_2^\dagger \hat{H}_N\hat{T}_2^2\rangle_L  + \mathcal{O}(t^5)+\dots
\label{eq:energy_functional_oddth_order_term}
\eeq
In the diagrammatic picture of Fig.~\ref{fig:dc_diagrams_open},
the $\mathcal{O}(t^3)$ part of the modified operator $\,\!{_2}\hat{T}_2$ can be viewed
as folding a pair of cluster operators, $\hat{T}_2^{\dagger}$ and $\hat{T}_2$,
into the original operator $\hat{T}_2$.
Higher-order contributions introduced by the transformation can then be interpreted as arising from repeatedly connecting additional cluster-operator pairs in the same way.

\subsection{QVDCD: DC modification to QVCCD}

\begin{table*}
\caption{\label{tab:comparison} Comparison of $\mathcal{O}(t^3)$ terms and diagrams
in energy functional of QVCCD and QVDCD.
}
\begin{ruledtabular}
\begin{tabular}{ccccc}
 &\multicolumn{2}{c}{QVCCD}&\multicolumn{2}{c}{QVDCD}\\
 label & diagram & $\mathcal{O}(t^3)$ in $2\langle\hat{H}(\,\!{_2}\hat{T}_2)\rangle$ &diagram & $\mathcal{O}(t^3)$ in $2\langle\hat{H}(\,\!{_2}\hat{T}_2)\rangle$ \\ 
\hline 
\rule{0pt}{30pt} A &
\begin{minipage}{0.1\textwidth}
\centering
\includegraphics[width=\linewidth]{diagrams/A.pdf}
{~~~$\mathbf{\bar A}$}
\end{minipage}
=
\begin{minipage}{0.1\textwidth}
\centering
\includegraphics[width=\linewidth]{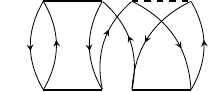}
{~~~$\mathbf{\bar A}_{ex}$}
\end{minipage}
 &\makecell{$-\frac{1}{4}(t^\dagger)^{ab}_{ij}\cdot (1-\tau_{ab})[(v_{kl}^{cd}-v_{lk}^{cd}) t_{ca}^{kl}t_{db}^{ij}]$\\ $=  -(t^\dagger)^{ab}_{ij} v_{kl}^{cd} t_{ca}^{kl}t_{db}^{ij}$}
 & 
\begin{minipage}{0.1\textwidth}
\centering
\includegraphics[width=\linewidth]{diagrams/A.pdf}
{~~~$\mathbf{\bar A}$}
\end{minipage}
 & $ -\frac{1}{2}(t^\dagger)^{ab}_{ij} v_{kl}^{cd} t_{ca}^{kl}t_{db}^{ij}$
 \\
 \rule{0pt}{30pt} B&
 \begin{minipage}{0.1\textwidth}
\centering
\includegraphics[width=\linewidth]{diagrams/B.pdf}
{~~~$\mathbf{\bar B}$}
\end{minipage}
=
\begin{minipage}{0.1\textwidth}
\centering
\includegraphics[width=\linewidth]{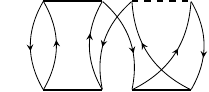}
{~~~$\mathbf{\bar B}_{ex}$}
\end{minipage}
 & \makecell{$-\frac{1}{4}(t^\dagger)^{ab}_{ij}(1-\tau_{ij})[(v_{kl}^{cd}-v_{lk}^{cd}) t_{cd}^{ki}t_{ab}^{lj}]$ \\ $=-(t^\dagger)^{ab}_{ij}v_{kl}^{cd}t_{cd}^{ki}t_{ab}^{lj} $}
 &\begin{minipage}{0.1\textwidth}
\centering
\includegraphics[width=\linewidth]{diagrams/B.pdf}
{~~~$\mathbf{\bar B}$}
\end{minipage}
 & $-\frac{1}{2}(t^\dagger)^{ab}_{ij}v_{kl}^{cd}t_{cd}^{ki}t_{ab}^{lj} $
 \\
 \rule{0pt}{30pt}C
 & 
\begin{minipage}{0.1\textwidth}
\centering
\includegraphics[width=\linewidth]{diagrams/C.pdf}
{~~~$\mathbf{\bar C}$}
\end{minipage}
=
\begin{minipage}{0.1\textwidth}
\centering
\includegraphics[width=\linewidth]{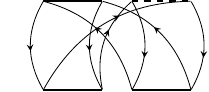}
{~~~$\mathbf{\bar C}_{ex}$}
\end{minipage}
 &\makecell{$\frac{1}{4}(t^\dagger)^{ab}_{ij}\cdot \frac{1}{2}(v_{kl}^{cd}-v_{lk}^{cd}) t_{cd}^{ij}t_{ab}^{kl}$ \\ $= \frac{1}{4}(t^\dagger)^{ab}_{ij}v_{kl}^{cd}t_{cd}^{ij}t_{ab}^{kl}$ }
 &-
 &-
 \\
 \rule{0pt}{30pt}D
 &
\begin{minipage}{0.1\textwidth}
\centering
\includegraphics[width=\linewidth]{diagrams/D.pdf}
{~~~$\mathbf{\bar D}$}
\end{minipage}
$\neq$
\begin{minipage}{0.1\textwidth}
\centering
\includegraphics[width=\linewidth]{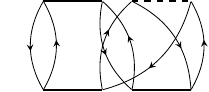}
{~~~$\mathbf{\bar D}_{ex}$}
\end{minipage}
 &\makecell{$\frac{1}{4}(t^\dagger)^{ab}_{ij}(1-\tau_{ab})(1-\tau_{ij})[(v_{kl}^{cd}-v_{lk}^{cd}) t_{ac}^{ik}t_{bd}^{jl}]$ \\ $=(t^\dagger)^{ab}_{ij}(v_{kl}^{cd}-v_{lk}^{cd})t_{ac}^{ik}t_{bd}^{jl}$}
 &\begin{minipage}{0.1\textwidth}
\centering
\includegraphics[width=\linewidth]{diagrams/D.pdf}
{~~~$\mathbf{\bar D}$}
\end{minipage}
 &$(t^\dagger)^{ab}_{ij}v_{kl}^{cd}t_{ac}^{ik}t_{bd}^{jl}$\\
\end{tabular}
\end{ruledtabular}
\end{table*}

The construction of QVDCD starts from the DC idea that, once the cluster excitation operator
is truncated, the fermionic exchange constraints associated with contractions involving
different excitation clusters may be relaxed, while the antisymmetry of the spin-orbital
doubles amplitudes is retained, without sacrificing the important properties of the theory 
(size-extensivity, exactness for $N$ particles, etc).
In the diagrammatic representation of the $\mathcal{O}(t^3)$ contributions to the QVCCD
functional in Table~\ref{tab:comparison},
replacing the antisymmetrized interaction $v^{kl}_{cd}-v^{lk}_{cd}$ by
the direct integral $v^{kl}_{cd}$ generates two diagrams for each contribution
to $\langle\hat{H}(\,\!{_2}\hat{T}_2)\rangle$, e.g., A is equivalent to the sum of $\mathbf{\bar A}$ and $\mathbf{\bar A}_{ex}$. 
Note: the diagrams in Table~\ref{tab:comparison} are not antisymmetrized diagrams.
Because the antisymmetry of the spin-orbital doubles amplitudes is retained,
diagrams $\mathbf{\bar A}$ and $\mathbf{\bar A}_{ex}$ give identical contributions, as do $\mathbf{\bar B}$ and $\mathbf{\bar B}_{ex}$, and $\mathbf{\bar C}$ and $\mathbf{\bar C}_{ex}$.
However, $\mathbf{\bar D}$ and $\mathbf{\bar D}_{ex}$ represent distinct contributions. 
The DC selection rule removes the quadruple contractions in which,
at each of the two vertices of the interaction line, the attached particle and hole lines
do not connect to the same double-excitation cluster, leaving only A, B, and $\mathbf{\bar D}$ in QVDCD.

A second requirement is that QVDCD remains exact with respect to CID in the isolated
two-electron and two-hole limits.
In those limits, the A, B, C, and D quadruple contributions
in QVCCD obey simple numerical relations, and $\mathbf{\bar D} = \mathbf{\bar D}_{ex}$.
In the two-electron limit, ${\rm A}+{\rm D}=0$, and $\frac{1}{2}{\rm B}+{\rm C}=0$.
In the two-hole limit,  ${\rm B}+{\rm D}=0$, and $\frac{1}{2}{\rm A}+{\rm C}=0$.
QVDCD is therefore constructed by removing the $\mathbf{\bar C}$, $\mathbf{\bar C}_{ex}$, and $\mathbf{\bar D}_{ex}$ contributions.
The contributions of A and B are further halved so as to recover the correct QVCCD limiting expressions.
The resulting $\mathcal{O}(t^3)$ contributions to the QVDCD functional are summarized in Table~\ref{tab:comparison}.

Note that in order to remove the exchange contributions in QVDCD,
the $\!{_2}t^{ij}_{ab}$ in the $\langle\hat{H}(\,\!{_2}\hat{T}_2)\rangle$ term
are contracted with the direct integral $v^{ab}_{ij}$ rather than the antisymmetrized integral,
\beq
\langle\hat{H}(\,\!{_2}\hat{T}_2)\rangle =& \frac{1}{2} v^{ab}_{ij}\,\!{_2}t^{ij}_{ab},
\label{eq:qvdcd_h_t}
\eeq
and so the symmetry of $\,\!{_q}t^{ij}_{ab}$ is relaxed,
i.e., $\,\!{_q}t^{ij}_{ab}$ in QVDCD only have the permutational symmetry
\beq
\,\!{_q}t^{ij}_{ab}=\frac{1}{2}(1+\tau_{ab}\tau_{ij})\,\!{_q}t^{ij}_{ab},
\eeq
and the renormalized double amplitudes are defined as
\beq
\,\!{_q}t^{ij}_{ab} &= \frac{1}{2}(1-\tau_{ab})(\,\!{_A}U^{-\frac{q}{2}})^c_at^{ij}_{cb} \\
&+ \frac{1}{2}(1-\tau_{ij})(\,\!{_B}U^{-\frac{q}{2}})_k^it^{kj}_{ab}\\
&- \frac{1}{2}(1+\tau_{ab}\tau_{ij}) (\,\!{_D}U^{-\frac{q}{2}})^{ci}_{ak}t^{kj}_{cb}.
\eeq

In the closed-shell spin-adapted form
\beq
\,\!{_q}\hat{T}_2 =& \frac{1}{2}\,\!{_q}T^{IJ}_{AB}\hat{E}^{A}_{I}\hat{E}^{B}_{J},\\
\langle\hat{H}(\,\!{_2}\hat{T}_2)\rangle=& \,\!{_2}T^{IJ}_{AB} (2v^{AB}_{IJ} - v^{BA}_{IJ}),\\
\langle\,\!{_1}\hat{T}_2^\dagger\hat{H}_N~\,\!{_1}\hat{T}_2\rangle =& \,\!{_1}T^{\dagger AB}_{~~IJ} \langle \Phi^{IJ}_{AB} \vert \hat{H}_{N} ~\,\!{_1}\hat{T}_2 \vert 0 \rangle,
\label{eq:closed_shell_form}
\eeq
where indices $A,B,C,\ldots$ on the right denote virtual spatial orbitals, 
$I,J,K,\ldots$ denote occupied spatial orbitals, and $P,Q$ denote any spatial orbitals.
The renormalized amplitudes in the spatial-orbital representation can be defined as
\beq
\,\!{_q}T^{IJ}_{AB}=\frac{1}{2}\Big(\frac{2}{3}+\frac{1}{3}\tau_{AB}\Big)&\Big(1+\tau_{AB}\tau_{IJ}\Big)
\Big[(\,\!{_A}U^{-\frac{q}{2}})_A^C\tilde{T}^{IJ}_{CB} \\
+&(\,\!{_B}U^{-\frac{q}{2}})_K^I\tilde{T}^{KJ}_{AB}
-(\,\!{_D}U^{-\frac{q}{2}})^{CI}_{KA}\tilde{T}^{KJ}_{CB}\Big],
\eeq
with
\beq
\tilde{T}^{IJ}_{AB} =& 2T^{IJ}_{AB} - T^{JI}_{AB}\\
\,\!{_A}U^A_B =& \delta^A_B + T_{~~IJ}^{\dagger AC}\tilde T^{IJ}_{BC}\\
\,\!{_B}U_I^J =& \delta_I^J + T_{~~IK}^{\dagger AB}\tilde T^{JK}_{AB}\\
\,\!{_D}U^{AJ}_{IB} =& \delta^A_B\delta_I^J + \tilde T_{~~IK}^{\dagger AC}\tilde T^{JK}_{BC} 
\label{eq:transformation_matrices}
\eeq

\subsection{Orbital optimization}
The orbitals are optimized by minimizing the energy functional with respect to orbital rotation
parameters,\cite{QVCC-J2}
\beq
\label{eq:orbital_gradient}
g^I_A =& F^I_A - T^I_JF^J_A - T^B_AF^I_B\\
& -T^J_K (2v^{IK}_{AJ} - v^{KI}_{AJ})  + T^C_B (2v^{IB}_{AC} - v^{BI}_{AC}) \\
& + \,\!{_1} T^{IK}_{CD} \,\!{_1} \tilde{T}^{CD}_{JL} v^{JL}_{AK} 
- \,\!{_1} {T}^{KL}_{AC}\,\!{_1} \tilde{T}^{BD}_{KL} v^{IC}_{BD}\\&
 - \frac{1}{2} (\,\!{_1} \tilde{T}^{IK}_{BD}\,\!{_1} \tilde{T}^{CD}_{JK} + 
3\,\!{_1} T^{KI}_{BD}\,\!{_1} T^{CD}_{KJ})v^{JB}_{AC}\\
& + \frac{1}{2} (\,\!{_1} \tilde{T}^{JL}_{AC}\,\!{_1} \tilde{T}^{BC}_{KL} + 3\,\!{_1} T^{LJ}_{AC}\,\!{_1} T^{BC}_{LK})v^{IK}_{BJ}\\
&+ \,\!{_1} \tilde{T}^{IK}_{BD} \,\!{_1} \tilde{T}^{CD}_{JK}v^{BJ}_{AC} - \,\!{_1} \tilde{T}^{JL}_{AC} \,\!{_1} \tilde{T}^{BC}_{KL}v^{IK}_{JB}\\
&+ \,\!{_2}\tilde{T}^{IJ}_{BC} v^{BC}_{AJ} - \,\!{_2}\tilde{T}^{JK}_{AB} v^{IB}_{JK},
\eeq
with
\beq
\,\!{_q}\tilde{T}^{IJ}_{AB} =& 2\,\!{_q} T^{IJ}_{AB} - \,\!{_q} T^{IJ}_{BA},\\
T^I_J =& \,\!{_1}\tilde{T}^{IK}_{AB}~ \,\!{_1}T^{AB}_{JK},\\
T^B_A =& \,\!{_1}\tilde{T}^{IJ}_{AC}~ \,\!{_1}T^{BC}_{IJ},\\
F^P_Q =& h^P_Q + (2v^{PI}_{QI} - v^{PI}_{IQ}),\\
\eeq
where $h^P_Q$ denote the one-electron integrals.
The orbital gradients, Eq.~(\ref{eq:orbital_gradient}), are used to update the rotation coefficients,
\beq
T^{I(n+1)}_A = T^{I(n)}_A+\frac{g^I_A}{F^I_I - F^A_A},
\eeq
which are then used to construct the unitary orbital rotation matrix,
\beq
\mathbf{U} = e^{\mathbf{T_1}-\mathbf{T_1^\dagger}}.
\eeq
The orbital rotation coefficients and doubles amplitudes are optimized simultaneously using 
the direct inversion of iterative subspaces (DIIS) acceleration
until the orbital gradients and amplitude residuals are reduced below the convergence thresholds.


\section{Results}
We implemented closed-shell QVCCD, QVDCD, and their orbital-optimized variants in \texttt{ElemCo.jl}.\cite{elemco} 
For small enough molecular systems, FCI serves as the benchmark.
For larger systems, we use internally contracted multireference CI with Davidson correction (MRCI+Q)\cite{MRCI,davidson}
or CCSD(T) (for weakly correlated systems) as the benchmark.
Both FCI and MRCI+Q calculations were performed in \texttt{MOLPRO},\cite{molpro}
whereas all remaining calculations were carried out in \texttt{ElemCo.jl}.
The RHF determinant was used as the reference,
core orbitals were frozen in the correlated calculations,
and density-fitting basis sets\cite{Density-fitting} were used throughout the \texttt{ElemCo.jl} calculations.

\subsection{Reaction energies}

We first assess OQVDCD in a regime dominated by dynamic correlation using 36 closed-shell
reaction energies from Ref.~\onlinecite{reaction_energy_set} in the aug-cc-pVTZ basis.\cite{cc-pVXZ-basis, aug-cc-pVXZ-basis}
The aug-cc-pVTZ JKFIT and MPFIT auxiliary bases are used for all atoms except Li,
for which the aug-def2-universal JKFIT basis\cite{def-df-basis} and the cc-pVTZ MPFIT basis
are employed.
We compare OQVDCD with CCSD, DCSD, OQVCCD, and CCSD(T),\cite{CCSD_T}
and summarize the deviations with the mean absolute deviation (MAD),
root-mean-square deviation (RMSD), and maximum deviation (MaxD).

\begin{table}
\caption{Reaction energy errors (kcal/mol) relative to CCSD(T) in aug-cc-pVTZ basis.}
\begin{ruledtabular}
\begin{tabular}{lccc}
Method & MAD & RMSD & MaxD \\
\hline
CCSD   & 1.08 & 1.67 & -5.56 \\
DCSD   & 0.85 & 1.24 & 4.41 \\
OQVCCD & 1.20 & 1.84 & -6.35 \\
OQVDCD & 0.81 & 1.18 & -3.21 \\
\end{tabular}
\end{ruledtabular}
\end{table}

For this reaction set, OQVDCD reduces the mean absolute deviation relative to OQVCCD
from 1.20 to 0.81 kcal/mol and the root-mean-square deviation from 1.84 to 1.18 kcal/mol,
bringing the errors close to DCSD.
This shows that the enhanced accuracy of the DC modification can be transferred to the QVCC framework.

\subsection{Bond dissociation of BeO}

\begin{figure}[ht]
\includegraphics[width=0.9\linewidth]{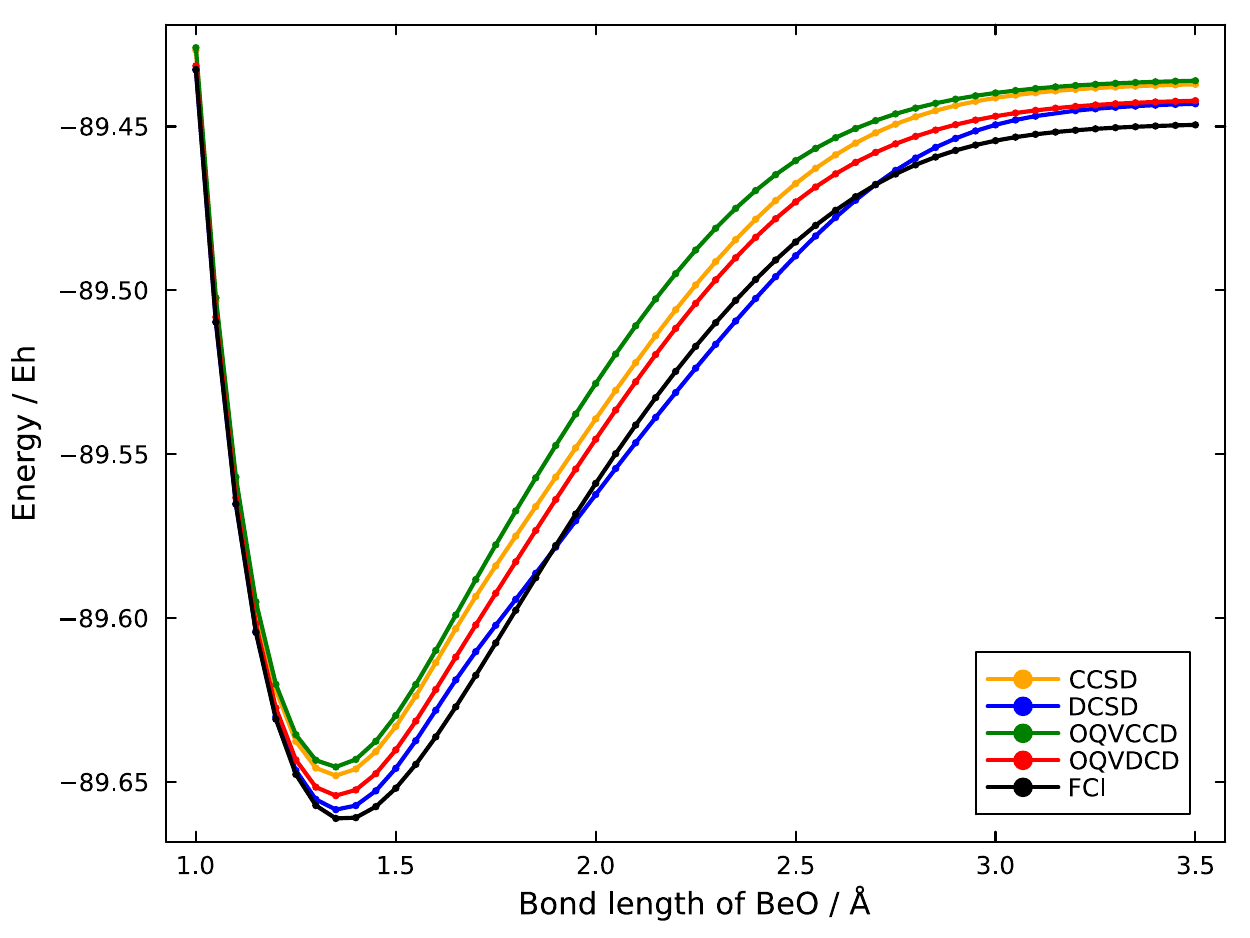}
\caption{Potential energy curves of BeO. CCSD, DCSD, OQVCCD, OQVDCD energies calculated with density-fitted cc-pVDZ basis, and FCI energies with cc-pVDZ basis.}
\end{figure}

\begin{figure}[ht]
\includegraphics[width=0.9\linewidth]{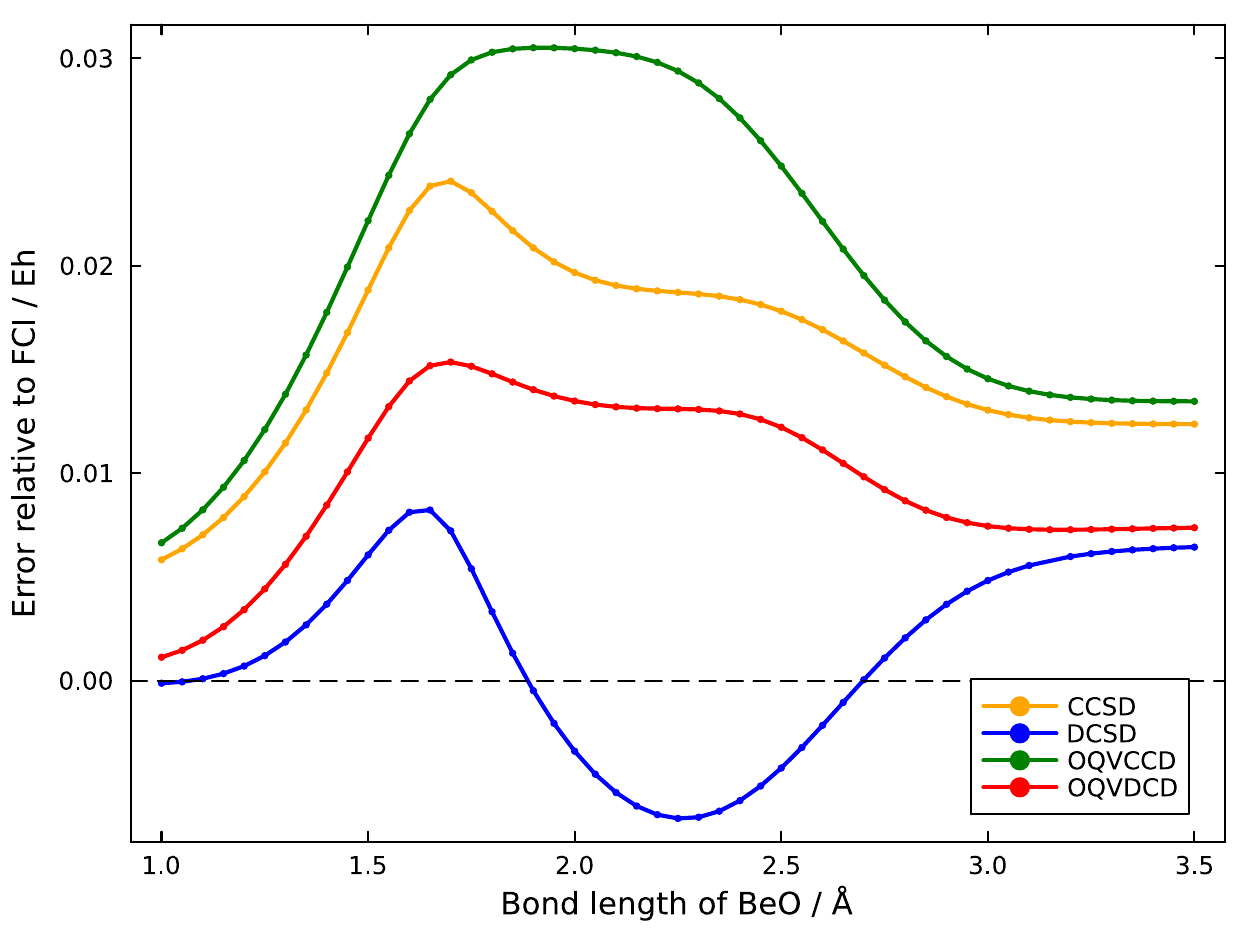}
\caption{Energy errors of CCSD, DCSD, OQVCCD, and OQVDCD for the BeO potential energy curves with respect to FCI.}
\end{figure}

We test the accuracy of the DC modification in bond dissociation,
where methods must remain reliable along the dissociation path as static correlation increases.
We have computed the BeO dissociation curve in the cc-pVDZ basis and compared it with FCI.
For DCSD, OQVCCD, and OQVDCD in \texttt{ElemCo.jl}, def2-universal JKFIT and the cc-pVTZ MPFIT
auxiliary basis sets were used.

In this example OQVCCD and OQVDCD behave variationally and remain above the FCI curve.
DCSD crosses the FCI curve even though its absolute deviation along the dissociation path is smaller.
OQVDCD reduces the OQVCCD deviation from FCI, consistent with the expected behavior
that the DC modification improves the description of dynamic correlation
while the quasi-variational nature of OQVDCD retains more upper-bound character than DCSD
in this system.
The non-parallelity error (NPE) of OQVDCD is 14.23 mEh, slightly smaller than that of DCSD,
14.87 mEh, whereas OQVCCD gives a larger value of 23.87 mEh.
If measured from the equilibrium geometry to the dissociation limit, the OQVDCD NPE is 8.40 mEh,
which is substantially smaller than the DCSD value of 14.87 mEh and the OQVCCD value of 17.05 mEh.

\subsection{Bond dissociation of N$_2$}
\begin{figure}[ht]
\includegraphics[width=0.9\linewidth]{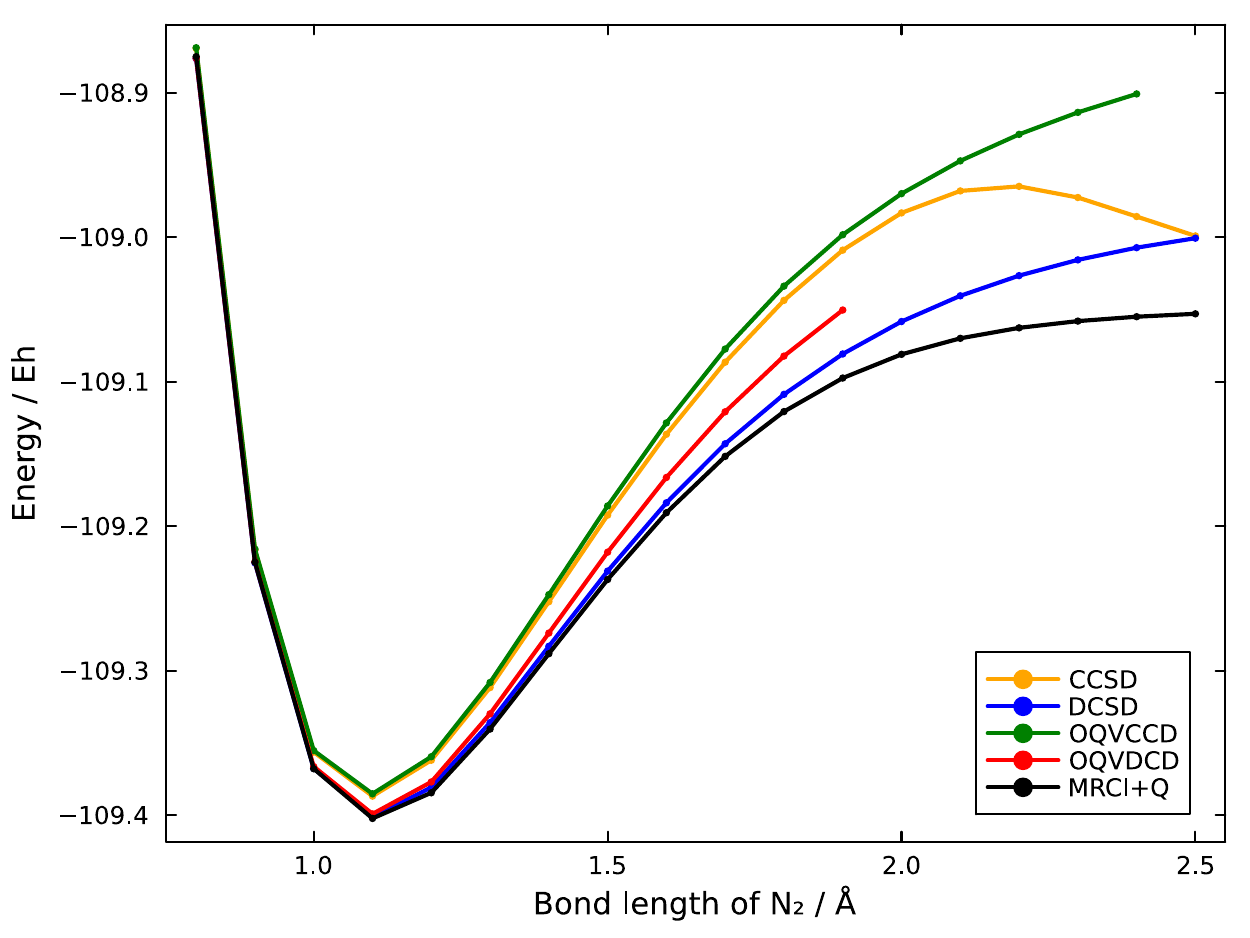}
\caption{Potential energy curves of N$_2$. CCSD, DCSD, OQVCCD, OQVDCD calculated with the density-fitted aug-cc-pVQZ basis, and MRCI+Q calculated with aug-cc-pVQZ basis.}
\end{figure}

N$_2$ potential energy curve is a standard strongly correlated test case for electron-correlation methods.
We calculated the potential energy curve in the aug-cc-pVQZ basis with the corresponding JKFIT and MPFIT
auxiliary basis sets.
Because FCI is not feasible here, MRCI+Q is used as the reference.

DCSD remains robust and qualitatively correct for this system,
whereas OQVCCD develops convergence problems at a bond length of 2.4 Å and beyond 
and also deviates more strongly from MRCI+Q.
This deviation can be reduced either by perturbative triples\cite{QVCC-J3} or,
by the DC modification, which is especially effective around equilibrium.
OQVDCD itself develops convergence problems beyond 1.9 Å.
Although it is less robust than DCSD in the strongly correlated regime,
it shifts the OQVCCD energies toward the benchmark and remains above the benchmark energies.

\subsection{H$_4$ rectangular distortion}

\begin{figure}
\includegraphics[width=0.9\linewidth]{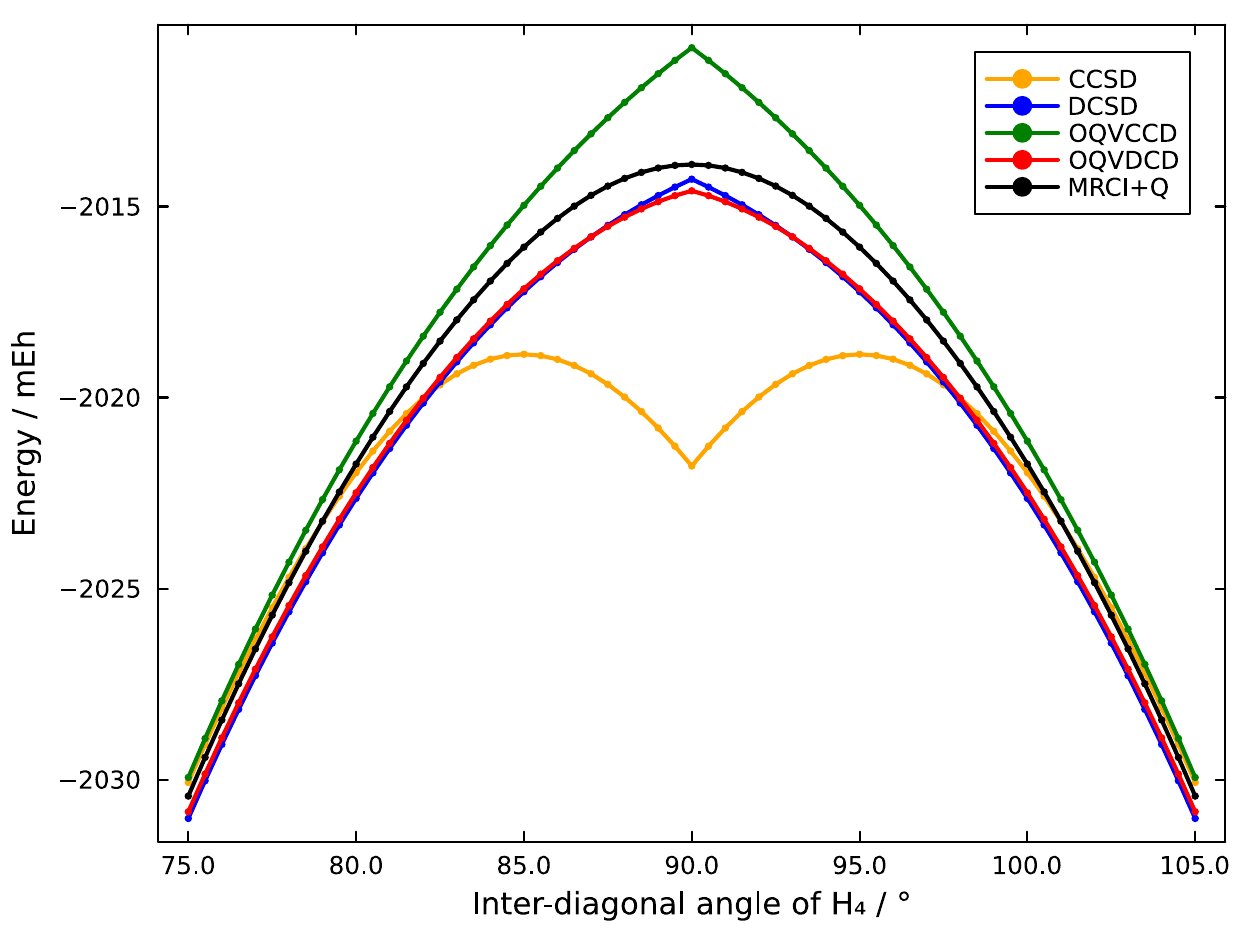}
\caption{\label{fig:H4} Potential energy curves of H$_4$ system. CCSD, DCSD, OQVCCD, OQVDCD calculated with the density fitted cc-pVQZ basis, and MRCI+Q calculated with cc-pVQZ basis.}
\end{figure}

A rectangular H$_4$ is a model that connects bond breaking with a metal-insulator-like transition.
The four H atoms sit on the vertices of a rectangle, each at a distance of 1.738 Å from the center.
As the angle between the diagonals changes from less than 90$^{\circ}$ to greater than 90$^{\circ}$,
the reference state switches between two pairing patterns. 
At the transition geometry of 90$^{\circ}$, single-configuration reference methods usually
exhibit an unphysical cusp (in the case of CCSD it even points down),
whereas the exact energy curve remains smooth.
We calculated the rectangular H$_4$ system in the cc-pVQZ basis with the corresponding JKFIT and MPFIT
auxiliary basis sets.

OQVCCD remains above the MRCI+Q curve but deviates from the benchmark the most.
DCSD curve is below the MRCI+Q curve but is more accurate overall.
OQVDCD behaves similarly to DCSD, but with a much smoother cusp at the transition.
The NPE of OQVDCD is 0.70 mEh, smaller than the DCSD value of 0.78 mEh
and substantially smaller than the OQVCCD value of 2.57 mEh.

\section{Conclusions}
In this work we presented OQVDCD by applying the DC approximation within the orbital-optimized
QVCC framework.
The central idea is to remove selected contributions from QVCCD that are 
associated with exchange between different double-excitation clusters in 
the Taylor-expanded QVCCD functional while retaining the quasi-variational structure,
the exact isolated two-electron and two-hole limits, and $\mathcal{O}(N^6)$ scaling.

On the reaction-energy test set, OQVDCD reduces the mean absolute deviation from CCSD(T) 
relative to OQVCCD from 1.20 to 0.81 kcal/mol and the root-mean-square deviation
from 1.84 to 1.18 kcal/mol, bringing the accuracy close to DCSD (and even slightly better).
In the strong-correlation tests considered here, OQVDCD generally shifts the OQVCCD energies
toward the benchmark, but its convergence is less robust than DCSD.
For the BeO and H$_4$ systems, OQVDCD reduces the non-parallelity error relative to OQVCCD and DCSD,
and for the rectangular H$_4$ system, it also smooths the cusp at the transition geometry.  
Nevertheless, these results suggest that the method is presently best suited to problems dominated
by dynamic correlation while still benefiting from the formal features of the quasi-variational framework,
such as the enhanced upper-bound character and the ability to define hermitian density matrices.

Several directions appear promising for future work.
Because OQVDCD retains a hermitian-like functional and well-defined density matrices, it is
a natural candidate for use as a solver within a multiconfigurational self-consistent field
(MCSCF) framework, where it could be employed to treat large active spaces (or molecular fragments)
that are inaccessible to a full configuration-interaction solver.
Such an optimization of the fragment is expected to improve the
insensitivity of the fragment energies to the orbitals, since it would provide a stationary
point of this dependency.\cite{Head-Gordon1998}
A further attractive avenue is the combination with linear-response and equation-of-motion
techniques. EOM-DCSD has been shown to be very accurate for excited states,\cite{rishiExcited2017,ElemcoilPaper}
and the variational (or unitary) coupled-cluster formulation offers a particular advantage in
linear-response theory: the matrix to be diagonalized is hermitian, so that complex eigenvalues,
which spuriously arise near conical intersections in non-hermitian formulations, cannot occur.
\cite{katsSecondorder2011,walzApplication2012}
Finally, the availability of a hermitian functional and density matrices makes the development
of analytical nuclear gradients and geometry optimization especially appealing, since DCSD is
already extremely accurate for equilibrium geometries and vibrational frequencies.
\cite{D3-Kats2015,kesharwaniSurprising2017,zieglerLocalized2019}
More broadly, the present method constitutes a first step toward the exploration of fully
hermitian distinguishable-cluster methods.


\begin{acknowledgments}
This work was supported by the Max Planck Society.
\end{acknowledgments}



\nocite{*}
\bibliography{aipsamp}

\end{document}